\documentstyle[12pt,epsfig]{article}
\input amssym.def
\input amssym.tex
\begin{document}
\begin{titlepage}
\title{Antiproton annihilation on light nuclei at very
low energies}

\author{K.V. Protasov \\
\small
Institut des Sciences Nucl\'eaires, IN2P3-CNRS, UJFG, \\ 
\normalsize
53, Avenue des Martyrs, 38026 Grenoble, Cedex, France\\ 
\large
G. Bonomi, E. Lodi Rizzini, A. Zenoni\\
\small
Dipartimento di Chimica e Fisica per l'Ingegneria e per i Materiali\\
\normalsize
Universit\`a di Brescia and INFN Sez. di Pavia, \\
\normalsize
Via Valotti 9, 25123 Brescia, Italy
}

\maketitle
\thispagestyle{empty}

\begin{abstract}
\normalsize
The recent experimental data obtained by the OBELIX group 
on $\bar{\mbox{p}}$D and $\bar{\mbox{p}}^4$He total annihilation cross
sections are analyzed. The combined analysis of these data with 
existing antiprotonic atom data allows, for the first time,
the imaginary parts of the S-wave scattering lengths for the two nuclei to be
extracted.
The obtained values are:
$\mbox {Im } a^{sc}_0 = [- 0.62 \pm 0.02 (\mbox{stat}) \pm 0.04
(\mbox{sys})] 
\mbox { fm}$ for $\bar{\mbox{p}}$D
and 
$\mbox {Im } a^{sc}_0 = [- 0.36\pm
0.03(\mbox{stat})^{+0.19}_{-0.11}(\mbox{sys})]
\mbox { fm}$
for $\bar{\mbox{p}}\:^4$He.
This analysis indicates an unexpected behaviour of
the imaginary part of the $\bar{\mbox{p}}$-nucleus S-wave scattering 
length as a function of the atomic weight $A$: \\
\begin{center}
$|\mbox {Im } a^{sc}_0|$ ($\bar{\mbox{p}}$p) $>$
$|\mbox {Im }~a^{sc}_0|$~($\bar{\mbox{p}}$D)
$>$ $|\mbox {Im }~a^{sc}_0|$~($\bar{\mbox{p}}^4$He). 
\end{center}
\end{abstract}

\bigskip

PACS numbers: 25.43.+t, 36.10$-$k  

\bigskip

{\it Submitted for publication to the {\bf European Physical Journal A}}
\end{titlepage}

\setcounter{figure}{0}

\section{Introduction}

Recent data obtained in the experiments on $\bar{\mbox{p}}$D and 
$\bar{\mbox{p}}^4$He
annihilation in flight at low energy~\cite{Zenoni99} and on the measurement 
of the shift and the width of the 1$S$ level of the $\bar{\mbox{p}}$D
atom \cite{Augsburger99} gave a first indication of a quite unusual
phenomenon. Indeed, the imaginary
part of the $\bar{\mbox{p}}$D scattering length appears to be
smaller than the $\bar{\mbox{p}}$p one~\cite{Protasov}.
The performed experiments
were the first investigations of the $\bar{\mbox{p}}$D system at
low energy; their realization was a very difficult task and the
precision and accuracy obtained were limited.

The first aim of this work is to combine the data coming from
these experiments in order to extract improved information on 
the imaginary part of the $\bar{\mbox{p}}$D scattering length. 
As we have shown in
our previous paper~\cite{Carbonell97}, devoted to the $\bar{\mbox{p}}$p
annihilation at low energies~\cite{Bertin96a,Benedettini97},
a successful analysis can
be done within the scattering length approximation written
for the systems with Coulomb attraction. These
two approaches, in-flight annihilation and antiprotonic atom experiments,
were shown to give coherent results and can be used as
complementary.
The second aim of this work is to perform a first phenomenological
analysis of the antiproton annihilation on heavier nuclei in S-wave
and in P-wave. 

The values of the $\bar{\mbox{p}}$D and $\bar{\mbox{p}}^4$He annihilation cross
sections used in this paper are reported in Tab.~\ref{t:tab1}. 
These are the
only $\bar{\mbox{p}}$-nucleus annihilation data available at incident 
momentum below 200~MeV/c. The data concerning shifts and widths of antiprotonic
atoms will be reported in the text.

\begin{table}[h!]
\centering
\caption{ Values of the $\bar{\mbox{p}}$D and $\bar{\mbox{p}}^4$He
          total annihilation cross sections used in this work,
          multiplied by the square of the incoming beam velocity $\beta$
	  for different $\bar{\mbox{p}}$ incident momenta. 
          In the data from~\cite{Zenoni99}, in addition to the quoted 
	  statistical and systematic
          errors, an overall normalization error of 2.5\% has to be considered.
          }
\vspace{0.5cm}
\begin{tabular}{| c| c |c |} 
\hline
     gaseous    &  $\bar{p}$ incident &  $\beta^2\sigma^T_{ann}$   \\
     target     &  momentum           &      (mbarn)             \\
      (ref.)    &  (MeV/c)            &              \\
\hline
D$_2$~\cite{Zenoni99} & 69.6$\pm$1.5   &  3.45$\pm$0.08 (stat) $\pm$0.15 (sys) \\
                    & 45.7$\pm$3.5   &  2.12$\pm$0.06 (stat) $\pm$0.33 (sys) \\
                    & 36.3$\pm$5.1   &  1.96$\pm$0.08 (stat) $\pm$0.55 (sys) \\
\hline
$^4$He~\cite{Zenoni99} & 70.4$\pm$1.3  &  4.63$\pm$0.10 (stat)$\pm$0.19 (sys) \\
                     & 47.0$\pm$3.3  &  2.45$\pm$0.10 (stat)$\pm$0.35 (sys) \\
\hline
 $^4$He~\cite{Balestra89} & 45.0$\pm$5.0  &  3.1$\pm$0.7 \\

\hline
\end{tabular}
\label{t:tab1}
\end{table}

The structure of the article is the following. In section~2
we present all the necessary scattering length formalism for systems
with Coulomb attraction. In section~3 the last experimental data on the
$\bar{\mbox{p}}$p annihilation cross sections are analyzed and the 
$\bar{\mbox{p}}$p low energy
parameters, extracted both from in-flight annihilation and from
atomic data, are shown to be in excellent agreement. Section~4 is
devoted to the combined analysis of the $\bar{\mbox{p}}$D data coming from
in flight annihilation and atomic experiments;
in this section the imaginary part of the $\bar{\mbox{p}}$D 
scattering length is obtained.
In section~5 we apply the analogous procedure to the $\bar{\mbox{p}}^4$He
annihilation data and we obtain the imaginary part of the $\bar{\mbox{p}}^4$He
scattering length. 
In addition, we examine the behaviour of 
$\bar{\mbox{p}}$-nucleus scattering parameters
as a function of the atomic weight $A$, for S-wave and P-wave, for
different nuclei.
Finally, the conclusions contain a brief summary of the results.

\section{Scattering length approximation}

The starting point to develop the scattering length approximation is
the relation between the $K$-matrix
for a given orbital momentum $l$ and the strong interaction phase shift
in presence of Coulomb forces
$\delta_l^{sc}$ \cite{MP}:
\begin{eqnarray}
\label{mainform}
\frac{1}{K_l^{sc} (q^2)} = g_l (\eta) q^{2l +1} [C_0^2(\eta)
\mbox{cot} \delta_l^{sc} - 2 \eta h(\eta)],
\end{eqnarray}
where $q$ is the center-of-mass momentum, $B$ the Bohr radius, $\eta = 1/qB$
\begin{eqnarray*}
g_0 (\eta) &=& 1, \nonumber\\
g_l(\eta) &=& \prod_{m=1}^l \left(1 + \frac{\eta^2}{m^2} \right),
\hspace{1cm} l= 1, 2, \ldots
\end{eqnarray*}
\begin{eqnarray*}
C_{0}^2(\eta)  =  {2\pi \eta \over1- \exp(-2\pi \eta)}; \hspace{1cm}
h(\eta)        = {1\over2}\left[\Psi(i \eta)+\Psi(-i \eta)\right]-
{1\over2}\ln\left(\eta^{2}\right)
\end{eqnarray*}
with the digamma function $\Psi$.
The $K$-matrix is related to the $S$-matrix by
\begin{eqnarray*}
S(q) = \frac{1 + ig_l(\eta)q^{2l+1} w(\eta)^*K}{1 - ig_l(\eta)q^{2l+1} w(\eta)K}
\end{eqnarray*}
with $w(\eta)=C_{0}^2(\eta)+2i\eta h(\eta)$.

The scattering length approximation used at low energies is equivalent to
the replacement of the $K$-matrix by a constant:
\begin{eqnarray*}
\frac{1}{K_l^{sc} (q^2)} = - \frac{1}{a_l^{sc}} +  o(q^2).
\end{eqnarray*}
\noindent Within this approximation, the annihilation cross section
$\sigma_{ann}^{l}$
for a given orbital momentum $l$ takes the form \cite{Carbonell93,Carbonell96}:
\begin{eqnarray}
q^2\sigma_{ann}^{l} =  (2l+1) 4\pi \,
\frac{g_l(\eta)q^{2l+1} \,C_0^2(\eta)\, {\mbox {Im }}(- a_l^{sc})}
{|1-ig_l(\eta)q^{2l+1} w(\eta) a_l^{sc} |^2}.
\end{eqnarray}

The scattering length approximation allows the
low energy parameters to be obtained also from atomic data. 
Following Trueman
\cite{Trueman}, it is necessary to replace in (\ref{mainform})
$q$ by $-i\sqrt{2\mu E}$, $\mu$ being the reduced mass,
$E= E_{nl} + \Delta E_{nl}$
the exact position of the Coulomb level $E_{nl}$ shifted and broadened by 
the strong interaction $\Delta E_{nl}$, and $\mbox{cot} \delta_l^{sc}$ by $i$:
\begin{eqnarray}
\frac{1}{a_l^{sc}} = - g_l (\eta) q^{2l +1} [C_0^2(\eta)
i - 2 \eta h(\eta)].
\end{eqnarray}
Within this  approximation (when one neglects effective
range corrections), this expression is exact. 
If one supposes that $\Delta E_{nl}/E_{nl} \ll 1$ or, equivalently,
$a_l^{sc}/B^{2l+1} \ll 1$ one obtains different approximate relations
usually called Deser \cite{Deser} (for S-wave in the first order) or
Trueman \cite{Trueman} (for S- and P-wave in higher orders) formulas.

\section{$\bar{\mbox{\bf p}}$p annihilation}

\begin{figure}[ht]
\begin{center}
\epsfysize=12cm
\centerline{\epsfbox{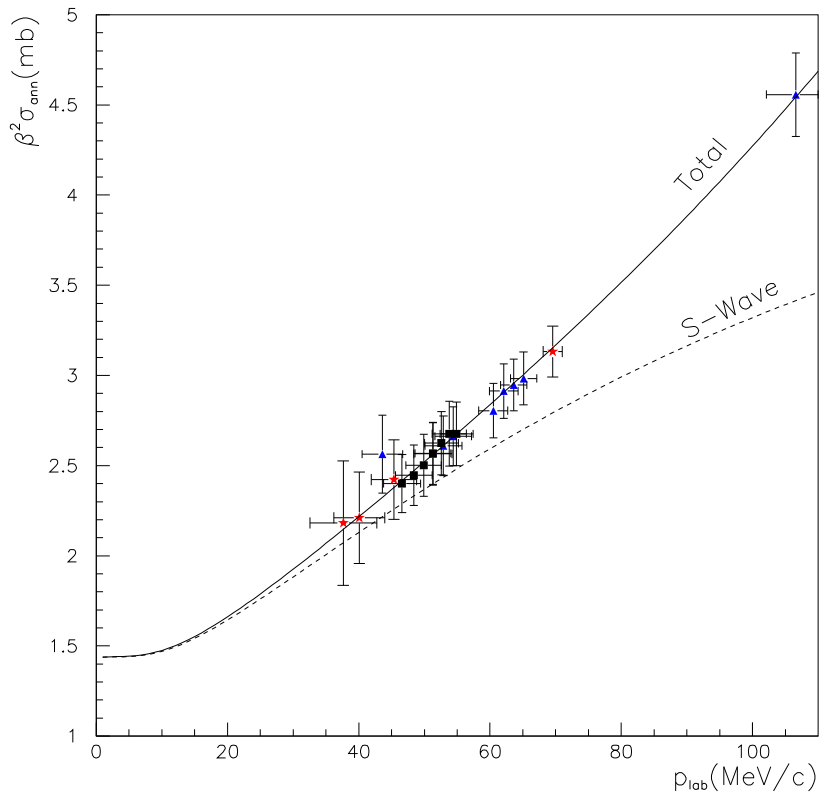}}
  \caption{Values of the total $\bar{\mbox{p}}$p annihilation cross section 
           multiplied by the square of the incoming beam velocity.
           Experimental data are from~\cite{Bertin96a}
	   ({\Large $\blacktriangle$}),
	   \cite{Benedettini97} ($\blacksquare$),~\cite{Zenoni992}
	   ({\Large $\star$}).
	   The error bars represent the quadratic addition of the statistical
	   error and of the systematic error interval divided by $\sqrt{12}$.
	   Moreover the data are affected by an overall normalization error:
	   3.4\% for the data from~\cite{Bertin96a,Benedettini97} and 2.5\%
	   for the data from~\cite{Zenoni992}.
           The theoretical curves are the result of the present work:
           the full line is the total annihilation cross section, the dashed 
           line represents the S-wave contribution.
   }
\label{pbarpfig}
\end{center}
\end{figure}

To illustrate the agreement between the values of the low energy
parameters extracted from the in-flight annihilation
and from the atomic data, let's start from
the $\bar{\mbox{p}}$p system, where the experimental data at very low energy 
are more abundant and precise.
This procedure was already applied to the $\bar{\mbox{p}}$p system
and is described in detail in \cite{Carbonell97}. In the present 
analysis, we add recently measured experimental points
\cite{Zenoni992}. The results of the fit are presented in Fig.~\ref{pbarpfig}.
As in the previous analysis~\cite{Carbonell97}, 
the experimental point at 43.6 MeV/c, which suffered from possible
unknown systematics~\cite{Zenoni992}, was not used for the fit. 
 
%

The fit was performed by means of the MINUIT program \cite{MINUIT} and 
provided the following best fit values for the imaginary parts
of the $\bar{\mbox{p}}$p scattering length (S-wave) and of the
scattering volume (P-wave):
\begin{eqnarray}
\label{firstfit}
\mbox {Im } a_0^{sc} = - [0.69 \pm 0.01 (\mbox{stat})
\pm 0.03 (\mbox{sys})] \mbox { fm}; \nonumber\\
\\
\mbox {Im } a_1^{sc} = - [0.75 \pm 0.05 (\mbox{stat})
\pm 0.04 (\mbox{sys})] \mbox { fm}^3 \nonumber
\end{eqnarray}
with the value of $\chi^2 = 0.25$ per point. In the fitting procedure only
the statistical errors were accounted for; their propagation produced 
the errors quoted as statistical in the values of the best fit parameters. 
The errors quoted as systematic come from the overall normalization error 
of the experimental data. 
Let us remind the reader that the following additional hypotheses
were used in the fitting procedure: the parameters are
the spin averaged ones,
$|\mbox { Re } a_0^{sc}| = |\mbox { Im } a_0^{sc}|$ and the real part of the
scattering volume can be neglected.

It is very instructive to compare the values obtained for the two parameters
with the ones obtained from the atomic data. 
The last world averaged value for the
shift and the width of the $1S$ atomic level of the $\bar{\mbox{p}}$p atom
is \cite{Gotta99}:
\begin{eqnarray*}
\Delta E_{1S} + i \frac{\Gamma_{1S}}{2} = [- (0.721 \pm 0.014) 
+ i(0.548 \pm 0.021)] \mbox { keV}
\end{eqnarray*}
which gives, by means of the Trueman formula, the imaginary part of 
the scattering length:
\begin{eqnarray*}
\mbox {Im } a_0^{sc} = - (0.694 \pm 0.027)  \mbox { fm},
\end{eqnarray*}
in excellent agreement with the result obtained from the in-flight
annihilation
data~(\ref{firstfit}).

The imaginary part of the P-wave scattering volume can be extracted
from the width of the $2P$ atomic level of the $\bar{\mbox{p}}$p atom. 
The majority of the atomic
experiments obtain this value indirectly through the intensity 
balance procedure which gives, actually, the lower limit for this
parameter (see discussion in~\cite{Carbonell97,Gotta99,Batty}).
The world averaged value
obtained by this method is~\cite{Augsburger992}:
\begin{eqnarray*}
\Gamma_{2P} = (32.5 \pm 2.1) \mbox { meV}
\end{eqnarray*}
which corresponds to:
\begin{eqnarray*}
\mbox {Im } a_1^{sc} = - (0.66 \pm 0.04)  \mbox { fm}^3.
\end{eqnarray*}
This value is smaller than the value obtained from the in-flight annihilation
data~(\ref{firstfit}).

However, the unique direct measurement of the width of the $2P$ atomic level,
which has been recently performed~\cite{Gotta99}, gives:
\begin{eqnarray*}
\Gamma_{2P} = (38.0 \pm 2.8) \mbox { meV}
\end{eqnarray*}
which corresponds to: 
\begin{eqnarray*}
\mbox {Im } a_1^{sc} = - (0.77 \pm 0.06)  \mbox { fm}^3
\end{eqnarray*}
in excellent agreement with (\ref{firstfit}).

The main conclusion of this analysis is that the results obtained from
these different experimental approaches are in quite good agreement;
therefore these experiments can be considered as complementary.

\section{$\bar{\mbox{\bf p}}$D annihilation}

The first measurement of the $1S$ level of the $\bar{\mbox{p}}$D atom
\cite{Augsburger99}
gave a very surprising result: the width of this level seems
to have the same size as the corresponding one of the $\bar{\mbox{p}}$p
atom.
The shift and the width of the $1S$ atomic level were found to be:
\begin{eqnarray*}
\Delta E_{1S} + i \frac{\Gamma_{1S}}{2} = [- (1.05 \pm 0.25) 
+ i(0.55 \pm 0.37)] \mbox { keV},
\end{eqnarray*}
which give the following scattering length for the $\bar{\mbox{p}}$D
system:
\begin{eqnarray*}
a_0^{sc} =[(0.7 \pm 0.2)  - i(0.4 \pm 0.3)  ]\mbox { fm}.
\end{eqnarray*}
Unfortunately, the experimental errors are too large to allow for an
unambiguous comparison with the $\bar{\mbox{p}}$p scattering length.

The precision of the imaginary part of the scattering length
can be improved significantly if
the information coming from atomic measurements are combined 
with the data on $\bar{\mbox{p}}$D
annihilation in flight. The general idea is quite simple:
the P-wave parameters are fixed from the atomic data (which
are quite precise in this case), the real part of the scattering
length is taken from the atomic data too (the annihilation
cross section is not very sensitive to this parameter). Thus
we can perform a fit to the annihilation cross section data with
only one free parameter: the imaginary part of the S-wave scattering
length.

The shift and the width of the $2P$ atomic level were measured 
directly with approximately ten percent accuracy \cite{Gotta99}:
\begin{eqnarray*}
\Delta E_{2P} + i \frac{\Gamma_{2P}}{2} = [- (243 \pm 26) 
+ i(245 \pm 15)] \mbox { meV}
\end{eqnarray*}
which correspond to the following scattering volume:
\begin{eqnarray*}
a_1^{sc} = [(3.3 \pm 0.3)  - i(3.18 \pm 0.18)] \mbox { fm}^3.
\end{eqnarray*}

\begin{figure}[ht]
\begin{center}
\epsfysize=10cm
\centerline{\epsfbox{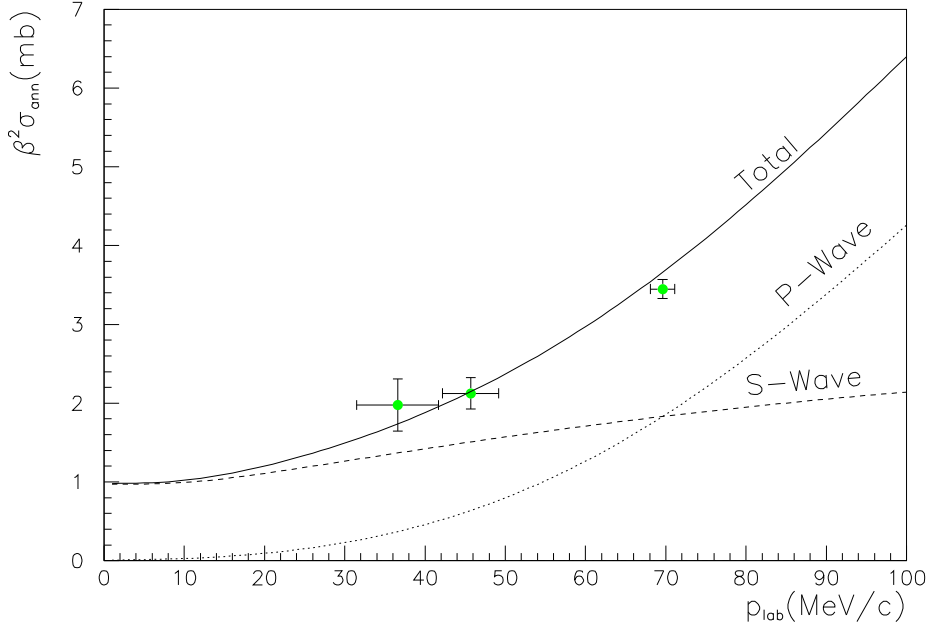}}
  \caption{
           Values of the total $\bar{\mbox{p}}$D annihilation cross section 
           multiplied by the square of the incoming beam velocity. 
           Experimental data are from~\cite{Zenoni99} ($\bullet$).
	   The error bars represent the quadratic addition of the
	   statistical error and of the systematic error multiplied
	   by $\sqrt{12}$. Moreover the data are affected by an overall
	   normalization error of 2.5\%.
           The theoretical curves are the result of the present work:
           the full line is the total annihilation cross section, the dashed 
           line represents the S-wave contribution, the dotted line the
	   P-wave contribution.
   }
\label{pbard}
\end{center}
\end{figure}
%
The results of the fitting procedure are presented in Fig.~\ref{pbard}; 
the procedure provided the following best fit value for the
imaginary part of the scattering length:
\begin{eqnarray*}
\mbox {Im } a_0^{sc} = - [0.62 \pm 0.02(\mbox{stat})
\pm 0.05 (\mbox{sys})] \mbox { fm}
\end{eqnarray*}
with a rather large value of the $\chi^2$, which amounts to 8.3 per
point. 

In the fitting procedure only the
statistical errors were accounted for; their propagation produced the error
quoted as statistical in the value of the best fit parameter. 
The quoted systematic error originates from two main sources of systematics.
The first one is the error on the determination of the imaginary part 
of the scattering volume (giving 0.03 fm) and the second one is connected 
to the overall normalization error of the data (giving 0.04 fm). 
These two errors are added quadratically.
On the contrary, the imaginary part of the scattering length is 
practically insensitive
to the variation of its real part, within the experimental
errors taken from the atomic experiment.

The agreement of the theoretical curve with the experimental data, within the
statistical errors, is not very good, as shown by the large value 
of the $\chi^2$.
This discrepancy could be due to at least two reasons. 
The experimental points, especially at the lowest values of the incident 
momentum, are affected by large individual systematic errors, 
due to the difficulty of separating annihilations coming from 
the different momentum components of the antiproton beam~\cite{Zenoni99}. 
Moreover, at the lowest value of the incident momentum, the spread of the
projectile momentum distribution is quite large (see Tab.~\ref{t:tab1});
as the annihilation cross section, in this momentum range, is rapidly
increasing with the lowering of the incident momentum,
the measured value of the cross section should be considered as an average 
value over the momentum interval, rather than the cross section value 
at the center of the interval.

In the error bars reported in Fig.~\ref{pbard} the two effects previously
mentioned are acconted for and the theoretical predictions appear in fair
agreement with the experimental data. Therefore, in spite of the large value of
the $\chi^2$ of the fit, we consider that the result obtained for the best
fit parameter could be confidently accepted. 

In conclusion, the combined analysis of the in-flight 
annihilation data and atomic data allows the knowledge of the imaginary part 
of the $\bar{\mbox{p}}$D scattering length to be improved significantly.
Moreover, the indication obtained from the atomic experiments that
this value does not exceed the $\bar{\mbox{p}}$p one is confirmed.

As a final comment, let us remark that, within a naive geometrical
approach to annihilation, one could expect
the imaginary part of the $\bar{\mbox{p}}$D scattering length
to be approximately equal to the sum of the $\bar{\mbox{p}}$n and
$\bar{\mbox{p}}$p scattering lengths:
\begin{eqnarray*}
\mbox {Im } a_0 (\bar{\mbox{p}}\mbox{D})\approx
\mbox {Im } a_0 (\bar{\mbox{p}}\mbox{n})+
\mbox {Im } a_0 (\bar{\mbox{p}}\mbox{p}).
\end{eqnarray*}
These results show that this vision is completely wrong especially
if one takes into account the quite
large value of the imaginary part of the $\bar{\mbox{n}}$p
(or equivalently $\bar{\mbox{p}}$n) scattering length
\begin{eqnarray*}
\mbox {Im } a_0 (\bar{\mbox{n}}\mbox{p})= -
[0.83 \pm 0.07 (\mbox{stat})]  \mbox { fm},
\end{eqnarray*}
which was obtained from the data on $\bar{\mbox{n}}$p annihilation
\cite{Mutchler}.

\section{$\bar{\bf p}^4$He annihilation and P-wave in different nuclei}

The ground state of the $\bar{\mbox{p}}^4$He atom is experimentally 
unaccessible.
Therefore one cannot obtain any information about the S-wave scattering
parameters from atomic experiments. Nevertheless, it is possible
to evaluate the imaginary part of the $\bar{\mbox{p}}^4$He scattering length
if one combines the atomic information with the data on in-flight
annihilation \cite{Zenoni99}. 

The P-wave scattering volume:
\begin{eqnarray*}
a_1^{sc} = [- (3.4 \pm 0.4) - i(4.4 \pm 0.5)] \mbox { fm}^3
\end{eqnarray*}
as well as the imaginary part of the
D-wave scattering parameter :
\begin{eqnarray*}
\mbox {Im } a_2^{sc} = - (1.42 \pm 0.06) \mbox { fm}^5
\end{eqnarray*}
can be extracted from the atomic data \cite{Schneider}.
Unfortunately, there is no experimental information about
$\bar{\mbox{p}}^4$He interaction in S-wave. To perform the fit,
we choose the value of the real part of the S-wave scattering length
$\mbox {Re } a_0^{sc} = 1.0$ fm, which is only slightly higher 
than the corresponding parameter for the 
$\bar{\mbox{p}}$p and $\bar{\mbox{p}}$D systems.
Given the large arbitrariness on this parameter, we 
assume an error of 50\% on its value ($\pm 0.5$ fm). 

\begin{figure}[ht]
\begin{center}
\epsfysize=10cm
\centerline{\epsfbox{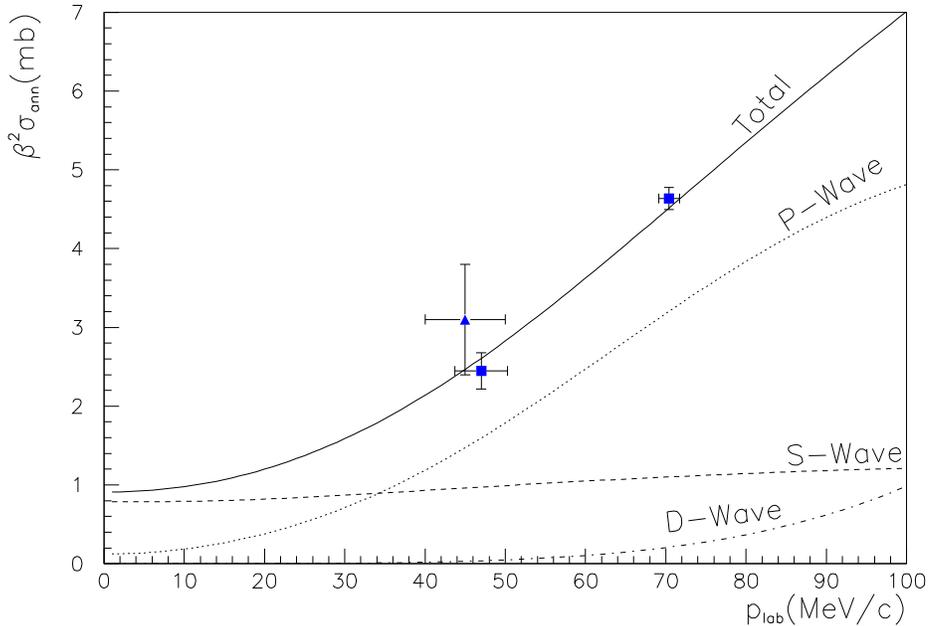}}
  \caption{Values of the total $\bar{p}^4$He annihilation cross section 
           multiplied by the square of the incoming beam velocity. 
           Experimental data are from~\cite{Zenoni99} ($\blacksquare$) 
	   and~\cite{Balestra89} ({\Large $\blacktriangle$}).
           The theoretical curves are the result of the present work:
           the full line is the total annihilation cross section, the dashed 
           line represents the S-wave contribution, the dotted line the
	     P-wave contribution,
	     the dashed-dotted line the D-wave contribution.
   }
\label{pbarHe}
\end{center}
\end{figure}

Thus, we can perform a fit of the available $\bar{\mbox{p}}^4$He annihilation 
cross section data with only one free parameter: the imaginary part of the
scattering length. The result of the fit is presented in
Fig.~\ref{pbarHe}; the experimental data considered are
from~\cite{Zenoni99} and from~\cite{Balestra89},
a previous low statistical measurement of the $\bar{\mbox{p}}^4$He annihilation 
cross section performed at LEAR with a streamer chamber.

%
The fit to the annihilation cross section data provided the following 
best fit value for the imaginary part of the scattering length:
\begin{eqnarray*}
\mbox {Im } a_0^{sc} = [-0.36 \pm 0.03 (\mbox{stat})
^{+0.19} _{-0.11} (\mbox{sys})] \mbox { fm}.
\end{eqnarray*}
with a value of $\chi^2$=1.8 per point.
In the fitting procedure only the statistical errors were accounted for; as in
the previous cases their propagation produced the error quoted as statistical in
the value of the fit parameter.
Here the systematic error contains two contributions. The first 
comes from the error on the imaginary part of the scattering volume 
and from the uncertainty on the real part of the scattering length; the
second comes 
from the overall normalization error of the data. 
The contribution coming from the imaginary part of the D-wave 
scattering parameter is negligible. 
As a final remark, let us emphasize that this is the first experimental 
evaluation of
the imaginary part of the $\bar{\mbox{p}}^4$He S-wave scattering length.

\begin{figure}[ht]
\begin{center}
\epsfxsize=10cm
\centerline{\epsfbox{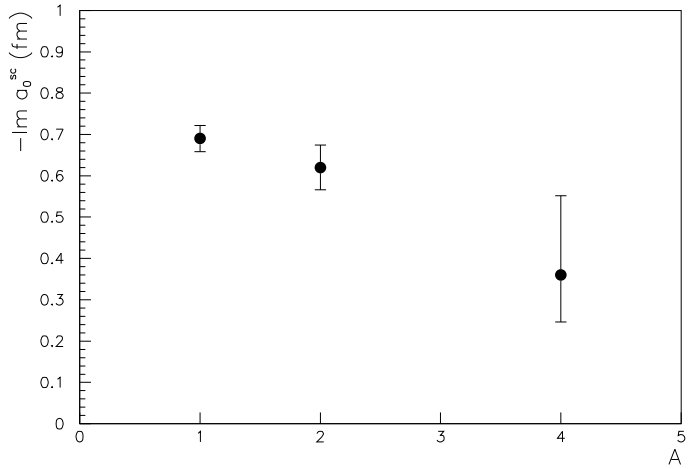}}
\caption{Absolute values of the imaginary part of the antiproton-nucleus
scattering length as a function of atomic weight $A$.
  Statistical and systematic errors are added quadratically.}
\label{recap}
\end{center}
\end{figure}
%
Figure ~\ref{recap} summarizes the values of the imaginary part of the
$\bar{\mbox{p}}$-nucleus 
scattering length as a function of the atomic weight $A$. 
This function seems to be a decreasing one.

\medskip

It is interesting to compare this unusual behaviour with the behaviour of the
imaginary part of the $\bar{\mbox{p}}$-nucleus  P-wave scattering volume 
as a function of the atomic weight.
For $\bar{\mbox{p}}$p, the information is obtained from
atomic and annihilation experiments, as described in section 3.
For heavier nuclei (D, He, Li) the data come from the
atomic experiments \cite{Gotta99,Schneider,Guigas},
where the shifts and the widths of the $2P$ levels
are measured directly. The experimental widths as well as
the imaginary parts of the scattering volumes, calculated through
the first order of the Trueman formula, are presented in
Tab.~\ref{TabVol}.  
\begin{table}[ht]
\centering
\caption{ $\bar{\mbox{p}}$-nucleus $2P$ level widths and Im~$a_1^{sc}$ 
calculated in the first order of the Trueman formula.}
\vskip 0.1 in                                                                   
\begin{tabular}{|c|c|c|c|} \hline                                  
  System     &ref.     & $\Gamma_{2p}$ & $-$Im~$a_1^{sc}$  \\ 
           &         &           (meV) &       (fm$^3$) \\
  \hline
$\bar{\mbox{p}}$H &\cite{Gotta99}    & $38.0\pm 2.8$         & $0.77\pm 0.06$ \\ 
$\bar{\mbox{p}}$D &\cite{Gotta99}    & $489\pm 30$           & $3.18\pm 0.20$\\
$\bar{\mbox{p}}^3$He&\cite{Schneider}& $(25\pm 9)\cdot 10^3$ & $3.1\pm 1.1$\\
$\bar{\mbox{p}}^4$He&\cite{Schneider}& $(45\pm 5)\cdot 10^3$ & $4.4\pm 0.5$\\
$\bar{\mbox{p}}^6$Li&\cite{Guigas}& $(444\pm 210)\cdot 10^3$ & $4.3\pm 2.0$\\
$\bar{\mbox{p}}^7$Li&\cite{Guigas}& $(456\pm 190)\cdot 10^3$ & $4.1\pm 1.7$\\ 
\hline
\end{tabular}
\label{TabVol}
\end{table}

\begin{figure}[ht]
\begin{center}
\epsfxsize=10cm
\centerline{\epsfbox{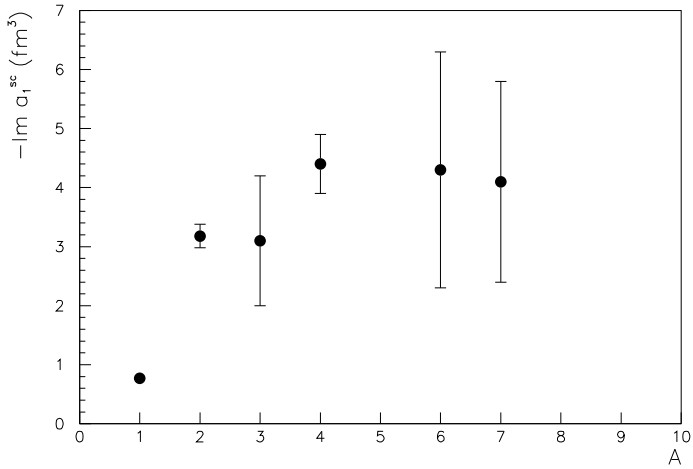}}
\caption{Absolute values of the imaginary part of the antiproton-nucleus
scattering volume as a function of the atomic weight $A$.
  }
\label{scvol}
\end{center}
\end{figure}

These values, as a function of the atomic weight, are presented
in Fig.~\ref{scvol}.
%
Unfortunately, the large experimental errors do not allow for
an unambiguous conclusion about the form of the functional 
dependence.
Nevertheless, this function seems to have, at least, a saturation-like
behaviour.

An analogous question can be put forward for the D-wave. 
The experimental atomic
information for heavy nuclei is quite abundant (see for review \cite{Batty95}).
However, the scattering length approximation used here to extract the 
scattering parameter $a_l^{sc}$ is no longer valid. When one works quite
far from the threshold,
it is necessary to take into account higher terms in the development
of the $K$-matrix and
 to use an effective range approximation instead of
the scattering length one. Thus, the $K$-matrix must be written as:   
\begin{eqnarray*}
\frac{1}{K_l^{sc} (q^2)} = - \frac{1}{a_l^{sc}} + 
\frac{1}{2} r_l^{sc} q^2 +  o(q^4).
\end{eqnarray*}
Here $q$ is the momentum of the corresponding Coulomb level:
$q=iq_B=i/nB$.
For light nuclei, like hydrogen or helium, for which B = 57.6
fm and B = 18 fm respectively, a correction
coming from the second term is negligible, if one supposes that
$r_l^{sc}$ is of the order of 1.0 fm. For heavier nuclei,
for instance $^{19}$F, $B = 3.37$ fm and this value is of the same order
as
all other parameters in this expression. Therefore, the development has no
sense and the scattering parameters cannot be extracted from the
results of atomic experiments. 
Note that the experiments measuring annihilation cross sections
have no such problem because, in principle, any
value of $q$ can be chosen. 

\section{Conclusions}

The proposed combined analysis of the recent $\bar{\mbox{p}}$-light nucleus
annihilation data and antiprotonic atom data allows our knowledge
of the low energy parameters in these systems to be improved.
The imaginary part of the spin-averaged $\bar{\mbox{p}}$D scattering
length can be obtained:
\begin{eqnarray*}
\mbox {Im } a_0^{sc} (\bar{\mbox{p}}\mbox{D})=
-[0.62\pm0.02(\mbox{stat})\pm0.05
(\mbox{sys)}] \mbox{ fm}.
\end{eqnarray*}
Moreover, for the first time, it becomes possible to evaluate
the imaginary part of the spin-averaged $\bar{\mbox{p}}^4$He scattering
length:
\begin{eqnarray*}
\mbox {Im } a_0^{sc} (\bar{\mbox{p}}^4\mbox{He})= [-0.36\pm0.03(\mbox{stat})
^{+0.19} _{-0.11}(\mbox{sys})]  \mbox { fm}.
\end{eqnarray*}
Compared to the 
the imaginary part of the spin-averaged $\bar{\mbox{p}}$p scattering
length
\begin{eqnarray*}
\mbox {Im } a_0^{sc} (\bar{\mbox{p}}\mbox{p})= -
[0.69 \pm 0.01 (\mbox{stat}) \pm 0.03 (\mbox{sys})]  \mbox { fm},
\end{eqnarray*}
obtained from annihilation data and in agreement with antiprotonic atom 
experiments, these values indicate the presence of a quite unexpected
phenomenon: the absolute value of the scattering
length seems to be a decreasing function of the atomic weight.

A naive geometrical vision of $\bar{\mbox{p}}$-nucleus
annihilation would suggest a value of the $\bar{\mbox{p}}$-nucleus 
scattering length increasing with the nucleus size. 
This naive picture is also wrong when one analyses the imaginary 
part of the P-wave scattering volume: the function
seems to have, at least, a saturation-like behaviour.

To confirm this unusual phenomenon involving low energy
$\bar{\mbox{p}}$-nucleus interactions, it would be necessary to perform
new measurements with higher statistics. This experimental information,
expanded to heavier nuclei, could be obtained with the new AD facility at
CERN.

\end{document}